\documentclass{article}
\usepackage{ijcai23}

\usepackage{times}
\usepackage{soul}
\usepackage{url}
\usepackage[hidelinks]{hyperref}
\usepackage[utf8]{inputenc}
\usepackage[small]{caption}
\usepackage{graphicx}
\usepackage{amsmath}
\usepackage{amsthm}
\usepackage{booktabs}
\usepackage{algorithm}
\usepackage{algorithmic}
\usepackage[switch]{lineno}
\usepackage[version=4]{mhchem}

\usepackage{multirow}

\usepackage{xcolor}

\usepackage{svg}
\usepackage{siunitx}
\usepackage{amssymb}
\newtheorem{definition}{Definition}

\newcommand{\Euc}{\mathrm{Euc}}
\newcommand{\M}{\mathcal{M}}
\newcommand{\R}{\mathbb{R}}

\newcommand{\Z}{\mathbb{Z}}
\newcommand{\ph}{\varphi}

\newcommand{\src}[1]{{\mathrm{src}(#1)}}
\newcommand{\tgt}[1]{{\mathrm{tgt}(#1)}}

\usepackage{multicol,multirow}
\usepackage{mathtools, stmaryrd}
\usepackage{xparse} \DeclarePairedDelimiterX{\Iintv}[1]{\llbracket}{\rrbracket}{\iintvargs{#1}}
\NewDocumentCommand{\iintvargs}{>{\SplitArgument{1}{,}}m}
{\iintvargsaux#1} %
\NewDocumentCommand{\iintvargsaux}{mm} {#1\mkern1.5mu,\mkern1.5mu#2}
\DeclareMathOperator*{\argmin}{arg\,min}
\usepackage{natbib}

\title{Unified Model for Crystalline Material Generation}

\author{
Astrid Klipfel$^{1,2,3}$
\and
Yaël Frégier$^3$\and
Adlane Sayede$^2$\And
Zied Bouraoui$^1$
\affiliations
$^1$Univ. Artois, UMR 8188, Centre de Recherche en Informatique de Lens (CRIL), F-62300 Lens, France.\\
$^2$Univ. Artois, UMR 8181, Unité de Catalyse et de Chimie du Solide (UCCS), F-62300 Lens, France.\\
$^3$Univ. Artois, UR 2462, Laboratoire de Mathématiques de Lens (LML), F-62300 Lens, France.\\
\emails
\{astrid.klipfel,yael.fregier,adlane.sayede,zied.bouraoui\}@univ-artois.fr,
}

\begin{document}

\maketitle

\begin{abstract}
One of the greatest challenges facing our society is the discovery of new innovative crystal materials with specific properties. Recently, the problem of generating crystal materials has received increasing attention, however, it remains unclear to what extent, or in what way, we can develop generative models that consider both the periodicity and equivalence geometric of crystal structures. To alleviate this issue, we propose two unified models that act at the same time on crystal lattice and atomic positions using periodic equivariant architectures. Our models are capable to learn any arbitrary crystal lattice deformation by lowering the total energy to reach thermodynamic stability. Code and data are available at \url{https://github.com/aklipf/GemsNet}. 
\end{abstract}

\section{Introduction}
One of the fundamental challenges of materials science is to obtain new thermodynamically stable materials that satisfy certain desirable properties. Recently, considerable attention has been devoted to the generation of crystalline (crystal) materials due to their wide range of usage in our modern society, e.g.\ metal alloys or semiconductors. As an example of application, we are interested in generating new crystal materials for developing new solar panels with a band gap enabling hydrolyse allowing hydrogen production from \ce{H2O}. Another application consists in discovering metal hydride for hydrogen storage applications which helps solve problems related to clean energy production and storage, which is one of the major challenges facing our society.

To discover new materials with desirable properties, high-throughput screening based on Machine Learning (ML) models is the most widely employed technique.  Despite several success stories, in particular, for organic molecule generations, more progress is required for crystal generation.  Crystals are three-dimensional periodic structures composed of a wide variety of chemical bonds and atoms which are often represented as a parallelepiped tiling, a.k.a crystal lattice or unit cell. The periodic structure of crystals makes it difficult to process when training generative models. Recently, graph-based representation models based on geometrically equivariant ML techniques have led to substantial performance increases across a wide range of supervised tasks such as property predictions and classification \cite{SchuttKFCTM17,jorgensen2018neural,gasteiger_dimenet_2020,gasteiger_dimenetpp_2020,doi:10.1021/acs.chemmater.9b01294,Choudhary2021,klicpera2022gemnet} and material generation \cite{satorras2021en,DBLP:conf/iclr/XieFGBJ22,Long2021,PhysRevMaterials.6.033801,https://doi.org/10.48550/arxiv.2202.13947} for crystals.  In \cite{DBLP:conf/aaai/klipfelPHFSB23}, a general framework that formulates an Equivariant Message Passing Neural Network (EMPNN) on periodic structures acting on crystal lattice without any label from the interaction forces and stress tensors has been proposed.  This model enforces a structuring bias for crystals using group actions with respect to the Euclidean $Euc(3)$ and $\text{SL}_3(\Z)$ groups by means of an equivariance property of Message Passing Neural Network (MPNN) layers.
Results on the denoising tasks show the capability of the EMPNN model to perform arbitrary crystal lattice deformation by improving the total energy of the structure (lowering the total energy of a structure leads to obtaining thermodynamic stability). In a similar vein, \cite{DBLP:conf/iclr/XieFGBJ22} proposed a model adapted from GemNet \cite{klicpera2022gemnet}, which is an equivariant GNN model to rotation and translation groups for organic chemistry, to periodic structures. Contrarily to the model from \cite{DBLP:conf/aaai/klipfelPHFSB23}, the model from \cite{DBLP:conf/iclr/XieFGBJ22} is capable to act on the atomic positions, but not capable of modifying the geometry (the shape of crystal lattice).

In this paper, we propose and analyse two unified models that process both the geometry and atomic positions of crystal materials at the same time. This is important for generation since the model will be able to modify the whole geometry during the inference and hence reach an optimum one. We first propose an Equivariant Graph Neural Network (EGNN) model as an extension of the EMPNN model that adds action on the atomic positions in addition to the shape of the lattice. While outperforming the EMPNN model, the architecture of EGNN remains suboptimal as 
it does not allow dealing with chemical structures. Namely, EGNN relies on a very basic embedding to represent angles whereas the message-passing schema only relies on the edges, but not the whole geometric information of the triplets.
To this end, we propose \textit{GemsNet}, a geometry architecture inspired by GemNet \cite{klicpera2022gemnet} that is capable to modify both atomic positions and lattice geometry at the same time within the same Equivariant GNN by exploiting all the geometric information of the triplets. To the best of our knowledge, \textit{GemsNet} and \textit{EGNN} are the first unified models that act simultaneously on crystal lattice geometry and properties. The quantitative and qualitative analyses of our models show the effectiveness of such a unified approach.

\section{Related Works}
Within the area of discovering new organic or crystal materials, we can distinguish different classes of related works according to representing technique used to encode molecules, e.g. Fingerprint representation \cite{doi:10.1021/acscentsci.0c00426,DBLP:journals/corr/abs-1810-11203} or voxel representation \cite{doi:10.1021/acs.jcim.0c00464,doi:10.1126/sciadv.aax9324,NOH20191370,Long2021}. From organic chemistry to material science, models based on graph representation are by far the most used in a variety of tasks such as classification \cite{SchuttKFCTM17,jorgensen2018neural,gasteiger_dimenet_2020,gasteiger_dimenetpp_2020,doi:10.1021/acs.chemmater.9b01294,Choudhary2021,klicpera2022gemnet} and generation \cite{satorras2021en,DBLP:conf/iclr/XieFGBJ22,Long2021,PhysRevMaterials.6.033801,https://doi.org/10.48550/arxiv.2202.13947}. 

The graph-based representation of materials has the advantage of representing the periodicity of structures as well as the local environment of each atom. The graph-based representation allows using Graphical Neural Networks (GNN) architectures to manipulate materials. GNNs are capable to process sparse data and take into account invariance or equivariance to many action groups that act on molecules to deform them. Existing works (e.g. \cite{klicpera2022gemnet}) are equivariant to $\text{SO}(3)$ due to a spherical basis which allows predicting lattice properties and performing simulations. However, this is not enough to deform crystal lattices when the shape of the lattice is unknown in advance. In \cite{DBLP:conf/aaai/klipfelPHFSB23}, group actions with respect to the Euclidean group $Euc(3)$ and $\text{SL}_3(\Z)$ group are incorporated by the equivariance property of MPNN layers to act on crystal lattices. This has the advantage of modifying the material regardless of its representation, i.e. the modification applied by the MPNN layer is independent of the orientation or the way the material is paved. Inspired by GemNet \cite{klicpera2022gemnet} an equivariant GNN model to rotation $SO(3)$ and translation groups for organic chemistry, \cite{DBLP:conf/iclr/XieFGBJ22} proposed a model for periodic structures that contrarily to \cite{DBLP:conf/aaai/klipfelPHFSB23} allows to act on atomic positions, but not capable of modifying the shape of crystal lattice.  While existing models have shown interesting results on denoising tasks, i.e.\ performing arbitrary crystal lattice deformation by improving the total energy of the structure, acting only on atomic positions or the crystal geometry is not enough to perform generation tasks. Our models perform simultaneously on both atomic positions and the crystal geometry within the same equivariant architecture.

Recently, several molecular dynamics models for generations have been proposed by approximating DFT simulation with GNN \cite{Pickard_2011,PhysRevMaterials.6.033801,https://doi.org/10.48550/arxiv.2202.13947,https://doi.org/10.48550/arxiv.2012.02920}. 
These methods mainly learn interaction forces and stress tensors to lower the total energy of a structure with methods analogue to DFT calculation. While relying on self-simulations to gather data is interesting, it only concerns a small set of specific structures. To discover new materials, however,  a lot of additional information about interaction forces is required, which is not always available. In addition, we cannot rely on randomly generated structures, as they lead in general to unstable structures. Our models do not require any additional labels.

\section{Background}


A crystalline (crystal) structure can be defined as a cloud of atoms and a repetition pattern. The repeated pattern represents periodicity and is often described as a parallelepiped called a cell. The repetition of the cell is called a lattice. The periodic structure is obtained with the tiling of the space by the crystal cell.  In particular, a given atom inside of the cell is repeated in multiple positions because of the tiling in space. As a consequence, the local environment of an atom can overlap with adjacent repetitions. We follow \cite{DBLP:conf/aaai/klipfelPHFSB23} to represent a crystal material as a graph. It is an oriented graph where each edge is represented by triplets containing the index of the source node, the index of the destination node and the relative cell coordinate of the destination node.  Notice that the definition of  the graph provided in \cite{DBLP:conf/aaai/klipfelPHFSB23} generalizes most of the existing graph definitions \cite{jorgensen2018neural,doi:10.1021/acs.chemmater.9b01294,satorras2021en}.

\begin{definition} 
  The representation space of {\em featured materials} ${\cal M}^F$ is the disjoint union $\coprod_{n \in \mathbb{N}} \M^F_n$ where:
    $$ {\cal M}^F_n =
        \big\{(\rho, x, z) \:|\:  \rho \in GL_d(\mathbb{R}),
        \: x \in [0, 1[^{n \times d},
        \: z \in F^n \big\}. $$
    Chemical materials are represented in $\M = \M^{\mathbb{N}}$, with atomic numbers as
    feature sequence $z$. 
\end{definition}

$\text{GL}_d(\mathbb{R})$ defines the shape of the lattice, i.e.\ the periodicity, where $d$ stands for the dimension of the material. $F$ is the feature space that encodes chemical information such as atomic number or charge.  Our aim is to  consider networks capable of deforming the geometry of a structure in order to minimize the total energy and hence obtain a stable structure.
To deform the geometry of a structure, we consider a neural network that predicts a collection $\{y, y_1, \cdots, y_n\}$, where $y\in\text{GL}_3(\mathbb{R})$ acts on the material lattice $\rho$ resulting in the updated lattice $\rho'$ where $y_i\in\mathbb{R}^3$ acts on the atomic positions as follows:
\begin{align*}
  f \colon {\cal M}_n^F &\to {\cal M}_n^F\\
  (\rho, x, z) &\mapsto (\rho', x', z)
\end{align*}
with
\begin{equation}
    \begin{cases}
        \rho' = y\rho       \\
        x'_i  = [x_i + y_i]. \\
    \end{cases}\,\label{deformation_network}
\end{equation}

The atomic positions are brought back into the crystal lattice by truncation. The equivariance property ensures that the actions performed on a given material are independent of any choice of representation. 
Namely, a given material has multiple equivalent associated representations, each of which depends on the ordering of the atoms, the orientation of the crystal and the tailing of the space. Our goal is then to process materials independently from their representations. To this end, we use group actions to determine the relationship between different representations. Group actions are the mathematical tools that enable one to pass from one choice of representation to another. The group $\Euc(3)$ of translations and rotations keeps track of the orientation and the position, while $\text{SL}_3(\Z)$ keeps track of the choice of a cell. This last group is needed because crystals are infinite collections of atoms while computers are only capable to fit a finite description. A natural approach to obtain such a finite description is to only work with a finite piece of a given crystal that is big enough to contain all the information of the structure. However, we need to ensure that the obtained results are independent of the choice of the piece of material we are working with. Since one can go from a description based on one choice of a piece of crystal to another using the $\text{SL}_3(\Z)$ action, results produced by models equivariant to this action will not depend on a particular choice of a representative piece of material.
More formally, equivariance is defined as follows:
$$ g\cdot f(M)=f(g\cdot M), g\in\Euc(3)\times SL_d(\Z) .$$
For more details about how an element $g$ of the group acts on the structure $M$, we refer to Proposition 1 and 2 in \cite{DBLP:conf/aaai/klipfelPHFSB23}. 

We now recall the EMPNN model proposed in \cite{DBLP:conf/aaai/klipfelPHFSB23} which performs arbitrary deformation by acting on relative atomic distances and angles. In this model, the spatial equivariance is enforced by an MPNN layer. Intuitively the EMPNN model takes advantage of the local invariance (input quantities are themselves invariant: distance, angle, etc.) and equivariance of the graph-based representation of materials to define equivariant actions on crystal lattices. 

\begin{definition}
\label{def6}
    Let $M = (\rho, x, z)$ in $\M_n^F$ be a material and $c_i > 0$
    for $1 \leq i \leq n$ denotes cutoff distances.
    We define a directed 2-graph
    $\Gamma = \Gamma_{M,c}$ by the graded components: 
    \begin{itemize} 
    \item $\Gamma_0 = \{1, \dots, n \}$ 
    \item $\Gamma_1 = \big\{ (i, j, \tau) \in \Gamma_0 \times \Gamma_0 \times
        \Z^d \: \big|\: || \rho (x_j - x_i + \tau) || < c_i \big\}$ 
    \item $\Gamma_2 = \big\{ (\gamma, \gamma') \in \Gamma_1 \times \Gamma_1
        \: \big|\: \src{\gamma} = \src{\gamma'} \big\}$
    \end{itemize}
\end{definition}

The following definition introduces the notations needed for our models.

\begin{definition}
    Let consider $M = (\rho, x, z) \in \M^F$ and $\Gamma = \Gamma_{M, c}$, we introduce the following 
    notations:
    \begin{itemize} 
    \item $e_{ij\tau}=(x_j - x_i + \tau)$ 
    for edge vector in lattice coordinates, 
    \item $v_{ij\tau} = \rho e_{ij\tau}$ for the edge vector 
    in physical space,
    \item $r_{ij\tau} = ||v_{ij\tau}||$ for the physical edge length, 
    \item $\theta_{ij\tau k\tau'}$ as the unoriented angle between $v_{ij\tau}$ and $v_{{ik\tau'}}$
    \end{itemize}
    Let us also write $e_\gamma, v_\gamma, r_\gamma, \theta_{\gamma\gamma'}$
    for the same quantities when we do not need to make vertices explicit.
    Note that $r_\gamma$ and $\theta_{\gamma\gamma'}$ are natural Euclid invariants. 
\end{definition}


\section{Unified Models for Crystal Material}
In this section, we propose two unified models that are capable to act on both atomic positions and crystal lattice. The first model called EGNN, which starts from the EMPNN model proposed in \cite{DBLP:conf/aaai/klipfelPHFSB23} which already acts on crystal lattices. The improvement consists in carefully adding a new component that modifies  atomic positions without losing the equivariance property. The other way around, we introduce a second model called \emph{GemsNet} which is based on GemNet \cite{klicpera2022gemnet,DBLP:conf/iclr/XieFGBJ22} which acts on atomic positions.  In GemsNet, we add components allowing to deform the crystal lattices within the same equivariant GNN architecture.  In the following, we describe these two models.

\subsection{Equivariant Graph Neural Network}

\begin{figure}
    \centering
    \includesvg[width=.41\textwidth]{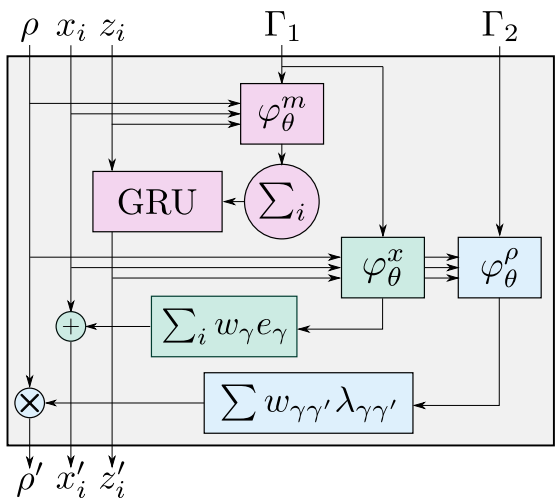}
    \caption{Our EGNN has a classic node-to-node message passing schema through the edges of the graph $\Gamma_1$ with the update of the chemical embedding by a GRU. In green colour, we have the update of atomic positions according to the chemical embedding of the atoms and the geometric information of the edges of the graph $\Gamma_1$. In blue colour, the generation of actions needed to be applied to the lattice from the information of the triplets and the edges is performed.}
    \label{fig:EGNN}
\end{figure}

The GNN model uses a message-passing schema with the formation of messages from the chemical and geometric information of the edges (Equation \ref{message_formation}) and updates the chemical features of the nodes with a GRU (Equation \ref{message_aggregation}). To generate a deformation of a crystal lattice, we first predict the weighting of the vector fields generated from the edges and the triplets (Equation \ref{weighting}). Then, we aggregate the weighting of all the vector fields to deform the lattice (Equation \ref{lattice_deformation}). Figure \ref{fig:EGNN} provides an overview of our EGNN model.

    \begin{equation}\label{message_formation}
     m^{l+1}_\gamma=\varphi^m_\theta(h^l_\src{\gamma}, h^l_\tgt{\gamma}, r^l_\gamma)
    \end{equation}
    \begin{equation}\label{message_aggregation}
        h^{l+1}_i=\textbf{GRU}(h^l_i, \sum_{\gamma\in\Gamma(i)} m^{l+1}_{\gamma})
    \end{equation}
    \begin{subequations}
        \begin{align}
        w^{l+1}_{\gamma} = & \varphi_\theta^{\rho^{(1)}}(h^{l+1}_\src{\gamma},h^{l+1}_\tgt{\gamma}, r^l_\gamma)\\
        w^{l+1}_{\gamma \gamma'} = & \varphi_\theta^{\rho^{(2)}}(h^{l+1}_\src{\gamma},h^{l+1}_\tgt{\gamma},h^{l+1}_\tgt{\gamma'},r^l_\gamma, r^l_{\gamma'}, \theta^l_{\gamma \gamma'})
        \end{align}\label{weighting}
    \end{subequations}
    \begin{equation}\label{lattice_deformation}
        \rho^{l+1} = \left(I_3+k(\sum_{\gamma \in \Gamma_1} w^{l+1}_{\gamma} \lambda_\gamma+\sum_{(\gamma, \gamma') \in \Gamma_2} w^{l+1}_{\gamma\gamma'} \lambda_{\gamma\gamma'})\right) \cdot \rho^l
    \end{equation}

The EGNN model modifies atomic positions by updating them according to the following equations:

\begin{equation}
    x'_i = [x_i + \frac{1}{|\mathcal{N}(i)|}\sum^L_{l=1}\sum_{j\in\mathcal{N}(i)}w^l_{ij\tau} e_{ij\tau}]
\end{equation}

\begin{equation}
    x^{l+1}_i = \big[x^l_i + \frac{1}{|\mathcal{N}(i)|}\sum_{j\in\mathcal{N}(i)}w^l_{ij\tau} e_{ij\tau}\big]
\end{equation}


\begin{equation}
    w^l_{ij\tau} = W'^l\text{silu}(W^l[h^l_i||h^l_j||emb(r_{ij\tau})]))
    \label{pred_wijtau}
\end{equation}

With $w^l_{ij\tau}$ are the weights at the layer $l$ for the set of edges $\mathcal{N}(i)$ connected to the atom $i$ and $e_{ij\tau }$ is the vector corresponding to the edge $(i, j, \tau)$. Intuitively, the GNN architecture predicts the atom displacement along the edges. We compute an average of these predictions. The atoms displacement is computed as the sum of the displacement predicted by all the layers of the network. We then obtain the weightings $w^l_{ij\tau}$ from a prediction on the chemical embedding of the atoms and the embedding of the interatomic distances according to  the setup given in  \cite{SchuttKFCTM17}, which is expressed by Equation  \ref{embedding_distance_schnet}. 
\begin{equation}
    emb(r_{ij\tau})_k=\exp({-c||r_{ij\tau}-\mu_k||^2})
    \label{embedding_distance_schnet}
\end{equation}
  
We use $c=10\si{\angstrom}$ and $\mu_k=k0.1\si{\angstrom}$ with $0\leq\mu_k<c$. Notice that the size of the hidden layer (from Equation \ref{pred_wijtau}) is 256.

\subsection{GemsNet}

\begin{figure*}
    \centering
    \includegraphics[width=\textwidth]{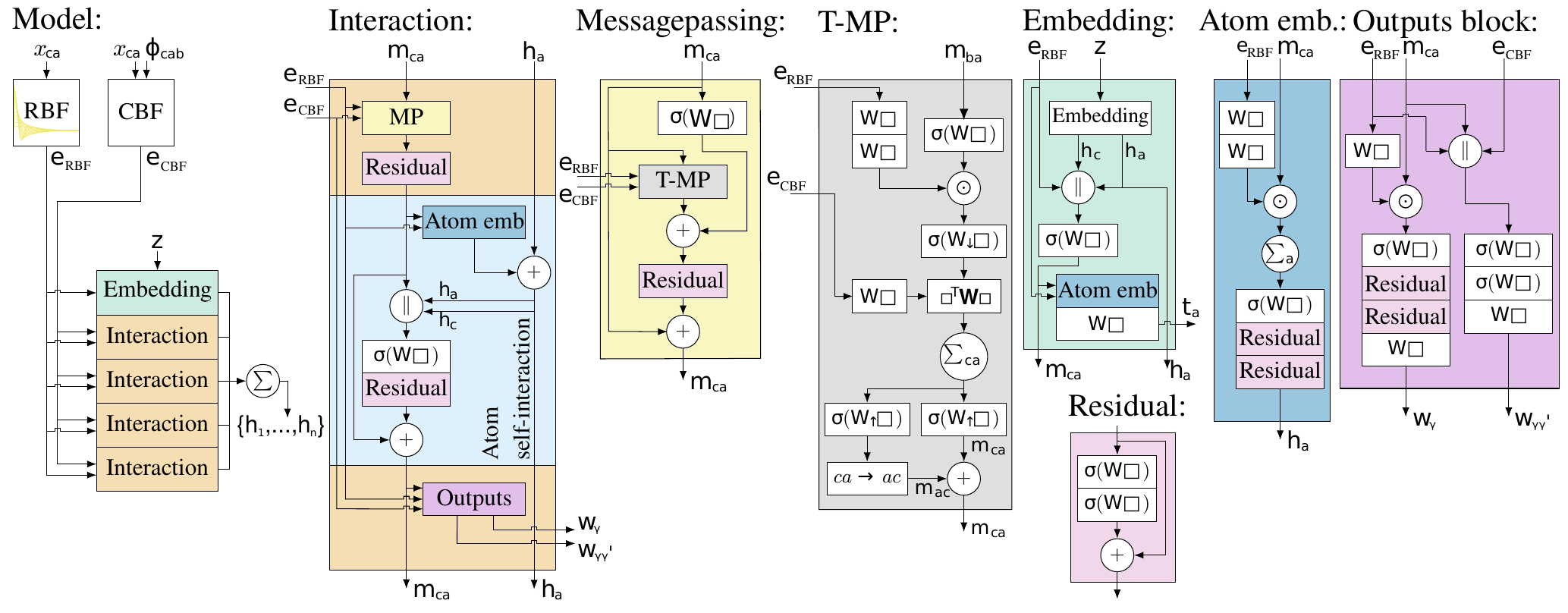}
    \caption{Overview of our GemsNet model. On the left, we have the whole network with the interaction layers stack that takes the geometric information as input. In orange colour, we can see the details about an interaction layer. We modify the last layer to predict atom displacements and lattice deformation. In yellow and grey, we have the diagram of message-passing between edges through triplets where the quadruplets have been removed from GemNet. On the right, we have the layers that update the chemical embedding of atoms. To update the atomic positions and the shape of the lattice, we aggregate the predictions of all the interaction layers.}
    \label{fig:gemsnet}
\end{figure*}


For GemsNet model, we start from the work of \cite{klicpera2022gemnet} on organic chemistry and its adaptation to crystal materials proposed by \cite{DBLP:conf/iclr/XieFGBJ22}. The overview of our proposed architecture is shown in Figure~\ref{fig:gemsnet}.
More precisely, we add a module to predict the actions on the lattice at the same time with atomic positions based only on triplets  from $\Gamma_2$. 
%
To perform crystal lattice deformation, we simply add components to the "output block" of GemNet to be able to predict deformations from the triplets based on the chemical embeddings. Recall that the "output block" is responsible for making predictions about the structures. Namely, it predicts the energy (when it is needed) and the interaction forces, which are used to generate a trajectory and to update the atomic positions). We add a feed-forward network to weight vector fields with $w^l_{\gamma\gamma'}\in\R^{N_\text{fields}}$ and obtain actions on the lattice.


\begin{equation}
\begin{aligned}
    w^l_{\gamma\gamma'} &= \ph^\rho_\theta(m^l_\gamma || m^l_{\gamma'} || emb_{\gamma\gamma'})\\
    &= W"^l\text{silu}\big(W'^l\text{silu}(W^l[m^l_\gamma || m^l_{\gamma'} || emb_{\gamma\gamma'}])\big)
    \label{pred_gamma_gammaprime}
\end{aligned}
\end{equation}

with $\ph^\rho_\theta$ is an MLP and $||$ is concatenation. $\theta_{\gamma\gamma'}$ is the concatenation of geometric features of GemNet triplet, so one RBF per edge and one CBF per angle such that: $emb_{\gamma\gamma'}=e_{\text{RBF}}(r_\gamma) || e_{\text{RBF}}(r_{\gamma'}) || e_{\text{CBF}}(r_\gamma,r_{\gamma\gamma'})$.

\begin{align}
    \tilde{e}_{\text{RBF},n}(r_\gamma)= & \sqrt{\frac{2}{c_\text{emb}}}\frac{\sin(\frac{n\pi}{c_\text{emb}}r_\gamma)}{r_\gamma}\\
    \tilde{e}_{\text{CBF},ln}(r_\gamma,\theta_{\gamma\gamma'})= & \sqrt{\frac{2}{c^3_\text{int}j^2_{l+1}}}j_l(\frac{z_{ln}}{c_\text{int}}r_\gamma)Y_{l0}(\theta_{\gamma\gamma'})
\end{align}

With the final representations of RBF and CBF are given as follows: 

\begin{align}
    e_{\text{RBF}}(r_\gamma)= & u(r_\gamma)\tilde{e}_{\text{RBF}}(r_\gamma)\\
    e_{\text{CBF}}(r_\gamma,\theta_{\gamma\gamma'})= & u(r_\gamma)\tilde{e}_{\text{CBF}}(r_\gamma,\theta_{\gamma\gamma'})
\end{align}

Where $u(r)$ is used to remove the interactions having distances greater than the cutoff distance. Following the recommendation given by \cite{gasteiger_dimenetpp_2020}, we chose $p=6$.
\begin{equation}
    u(r)= 1-\frac{(p+1)(p+2)}{2}r^p+p(p+1)r^{p+1}-\frac{p(p+1)}{2}r^{p+2}
\end{equation}

The hidden layers (Equation \ref{pred_gamma_gammaprime}) are of size 64. Notice that $m_{\gamma\gamma'}$ is invariant. As such, we can express $\rho'$ as follows: 
\begin{equation}
    \rho' = (I_3+\frac{1}{|\Gamma_2|}\sum^{L}_{l=1}\sum_{(\gamma,\gamma')\in\Gamma_2}{w^{l\intercal}_{\gamma\gamma'}}\lambda_{\gamma\gamma'})\rho
\end{equation}

with $\lambda_{\gamma\gamma'}\in\R^{N_\text{fields}\times 3\times 3}$ are the vector fields that define how the GNN acts on the crystal lattice. We use gradients of the geometric invariant such as: 

\begin{equation}
\lambda_{\gamma\gamma'}(M) = \frac{\partial r_\gamma}{\partial \rho} \oplus \frac{\partial r_{\gamma'}}{\partial \rho} \oplus \frac{\partial \theta_{\gamma\gamma'}}{\partial \rho}
\end{equation}


\section{Experiments}

\begin{table*}[t]
    \centering

    \begin{tabular}{lccccccccc}
        \toprule 
        \multirow{2}{*}{} & \multicolumn{3}{c}{Mp-20} & \multicolumn{3}{c}{Carbon-24} & \multicolumn{3}{c}{Perov-5}\\
        & lengths & angles & pos & lengths & angles & pos & lengths & angles & pos\\
        \cmidrule(lr){2-4}
        \cmidrule(lr){5-7} 
        \cmidrule(lr){8-10}
        EGNN & 0.345 & 4.394 & 0.039 & 0.631 & 9.546 & 0.068 & 0.348 & 4.450 & \textbf{0.028}\\
        GemsNet & \textbf{0.206} & \textbf{3.966} & \textbf{0.034} & \textbf{0.451} & \textbf{6.857} & \textbf{0.054} & \textbf{0.279} & \textbf{1.725} & 0.057\\
        \bottomrule  
    \end{tabular}
    \caption{We can see the reconstruction errors according to the model and the database. The angle and distance metrics are defined using Equations \ref{metrics_angles} and \ref{metrics_lengths}. They correspond to an average absolute error between the original and reconstructed mesh parameters in Angstrom for the distances and degrees for the angles. The atomic position metrics is an average distance between the original position and the reconstructed position as defined in Equation \ref{metrics_pos}}
    \label{table:benchmark}
\end{table*}


In this section, we experimentally compare the two proposed models with a number of baseline methods with respect to reconstruction and denoising tasks\footnote{Code and data are available at \url{https://github.com/aklipf/GemsNet}}. The denoising task (or geometry optimization task) aims to evaluate the capability of our model to perform arbitrary crystal lattice deformation by improving the total energy of the structure. To perform denoising, we apply a small random deformation to a stable structure leading to a less stable one with a high energy level (as energy increases in all directions locally). We can therefore generate pairs of stable and less stable structures that can be used to teach our models how to deform the less stable structure to reach stability. Making a structure more stable comes down to lowering the energy level (thermodynamic stability). The reconstruction task (or lattice reconstruction task) aims to restore a given crystal lattice from a non-noisy atomic coordinate following the same settings reported in \cite{DBLP:conf/aaai/klipfelPHFSB23}. More specifically, the reconstruction aims to build a crystal lattice from scratch. We start from the point cloud as if it was in a cubic lattice of one angstrom on a side. The main working hypothesis is that there is a single stable lattice which corresponds to the starting atomic positions. Notice that it is impossible to perform such a task with chemical simulation techniques such as DFT. Interestingly enough, with reconstruction capabilities, a model can be easily integrated into VAE or GANs architectures to perform generation.

\paragraph{Datasets.} We considered datasets of stable crystals where each structure is in local minima of formation energy. First, we use Perov-5 \cite{castelli2012new,castelli2012computational} which contains perovskite (cubic) structures with highly uniform shapes, but different chemical compositions between structures.  Second, we consider Carbon-24 \cite{carbon2020data} which is composed of carbon atoms having a large variety of shapes. Finally, we conduct experiments on Mp-20 \cite{jain2013commentary} which is a subset of the materials project proposed in \cite{DBLP:conf/iclr/XieFGBJ22} that has a large sample of shapes and chemical compositions. It is the most representative of ordinary structures. We used the same training, validation and test splits as reported in \cite{DBLP:conf/iclr/XieFGBJ22}.

\paragraph{Random deformation.} As mentioned above, the random deformation aims to introduce noisy structures needed to assess the capability of our models to reach stability. To this end, we sample noises and add them to the atomic positions. The lattice parameters are removed and replaced by a cubic lattice of dimension 1. The following equation gives precision about how to generate the noise.

\begin{equation}
\left\{ \begin{array}{ll}
\tilde{x_i} & = [x_i + \epsilon]\\
\tilde{\rho} & = I_3
\end{array} \right.,\epsilon\sim\mathcal{N}(0,0.05).\label{initial_state}
\end{equation}

\paragraph{Loss function.} As loss function, we aim to determine the optimal trajectory to denoise a structure. Indeed, as the crystals are periodic, several trajectories are possible to obtain the same result. The optimal trajectory is then defined as the trajectory that minimizes the distance we are moving through. More formally, we have: 

\begin{equation}
    y^*_i = x_i-\tilde{x_i}+\argmin_{\tau\in\mathbb{Z}^3} ||\rho(x_i-\tilde{x_i}+\tau)||
\end{equation}

We denote by $p:GL_3(\R)\to\R^6$ an application that associates the lattice parameters $(a, b, c, \alpha, \beta, \gamma)$ to the lattice $\rho$. We recall that $y_i$ is defined in \ref{deformation_network}. We can then define a reconstruction loss with L1 norm as follows:

\begin{align}
\mathcal{L}^\rho&=\frac{1}{6}\sum_{j=1}^6|p(\rho')-p(\rho)|_j\label{loss_params} \\
\mathcal{L}^x&=\frac{1}{3n}\sum^n_{i=1}\sum_{j=1}^3|y_i-y^*_i|_j \\
\mathcal{L}&=\mathcal{L}^\rho+\mathcal{L}^x
\end{align}

\paragraph{Evaluation metrics.} As evaluation metrics, we rely on angle and length metrics which are respectively Mean Absolute Error (MAE) metrics on the lattice parameters $(\alpha,\beta,\gamma)$ and $(a,b,c)$. The positioning metric is the distance between the optimal position and the actual position. We take into account the periodicity of the material for the atomic positions. If the atom is close to an edge and the optimal position is close to the opposite edge, then the distance to be covered is shorter than the distance that crosses the entire lattice. The distance metric is therefore the shortest distance between the position of an atom and one of its possible optimal positions. $N$ is the number of structures and $M$ the number of atoms.
\begin{align}
\text{lengths}&=\frac{1}{3N}\sum^N_{i=1}\sum^3_{j=1}|p(\rho_i')-p(\rho_i)|_j\label{metrics_lengths} \\
\text{angles}&=\frac{1}{3N}\sum^N_{i=1}\sum^6_{j=4}|p(\rho_i')-p(\rho_i)|_j\label{metrics_angles} \\
\text{pos}&=\frac{1}{M}\sum^M_{i=1}||y_i-y^*_i||\label{metrics_pos}
\end{align}

\paragraph{Training.} We train our two neural network models for 128 epochs. We use hard training from 45 min for the smallest dataset such as Perov-5 to 5 hours for the biggest datasets such as Mp-20. Our models are trained on an NVIDIA RTX A6000 GPU. The different hyperparameters are given in Table \ref{table:hyperparameters}. 

\begin{table}
    \centering
    \begin{tabular}{lcccc}
         \toprule 
        \multirow{2}{*}{} & \multicolumn{2}{c}{Carbon-24}  &\multicolumn{2}{c}{Mp-20} \\ 
        & lengths & angles & lengths & angles \\ 
        \cmidrule(lr){2-3}
        \cmidrule(lr){4-5}
        Baseline & 0.469 & 13.693 & 0.534 & 6.324\\
        EMPNN & 0.200 & 3.199 & 0.174 & \textbf{1.965} \\
        GemsNet & \textbf{0.176} & \textbf{3.109} & \textbf{0.130} & 3.082 \\
         \bottomrule  
    \end{tabular}
    \caption{MAE between lattice parameters of the original cell and the reconstructed cell as defined in Equations \ref{metrics_angles} and \ref{metrics_lengths}(in angstrom and in degree).}
    \label{table:reconstruction}
\end{table}

\begin{table}
    \centering
    \begin{tabular}{lcc}
        \toprule 
        Parameters & GemsNet & EGNN\\
        \midrule
        epoch & 128 & 128\\
        batch size & 256 & 128\\
        knn & 32 & 16\\
        lr & 0.001 & 0.0003\\
        layers & 3 & 6\\
        features & 256 & 128\\
         \bottomrule  
    \end{tabular}
    \caption{Hyperparameters for GemsNet and EGNN models.}
    \label{table:hyperparameters}
\end{table}


\paragraph{Lattice reconstruction task.} Following the same setting proposed in \cite{DBLP:conf/aaai/klipfelPHFSB23}, we use Equation \ref{initial_state} without applying noise to $x_i$, i.e.\ $\epsilon:=0$. We only use the loss term on the Equation \ref{loss_params} of lattice parameters. Table \ref{table:reconstruction} reports the obtained  results. We can see that GemsNet has better performances than existing models on Carbon-24 and Mp-20 datasets. In particular, there is an improvement in the length parameters. However, we obtain identical or slightly worse results for the angular parameters. The results are more variable on Mp-20 than on Carbon-24. This is expected since Carbon-24 is a database composed only of pure carbon structures under pressure, while Mp-20 contains many different chemical species and forms. It is therefore possible that the training is less stable on Mp-20 because there is more variability from one structure to another. Overall, GemsNet achieves good performance. This is interesting because GemsNet has more than 2 times fewer parameters and it is much faster in terms of inference compared to existing models.

\subsubsection{Geometry optimization task}
\begin{figure*}
    \centering
    \includegraphics[width=\textwidth]{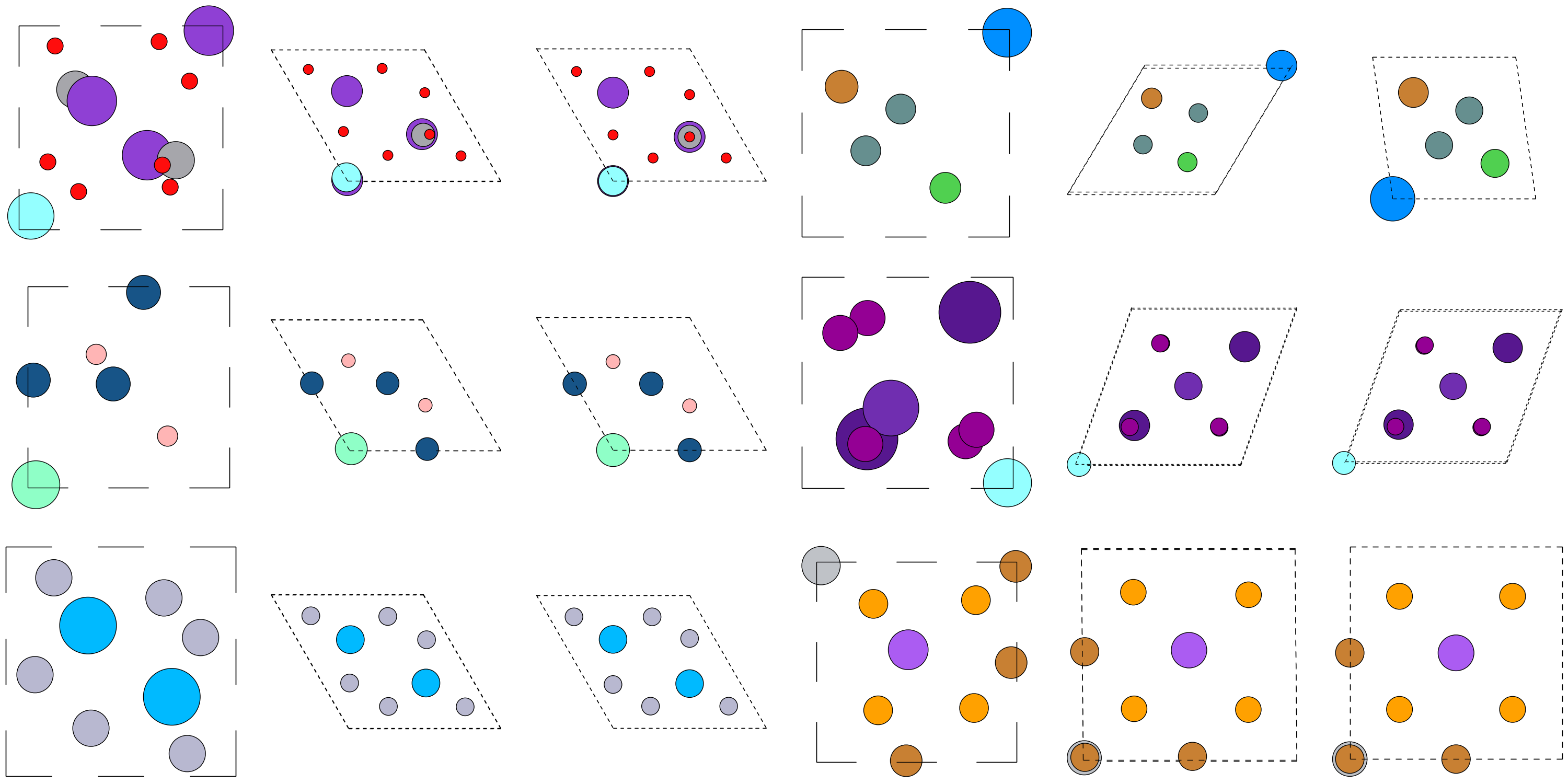}
    \caption{Action of GemsNet on 6 structures from the Materials Project. The left column is the noisy structure given as input to GemsNet. The central column is the structures updated by GemsNet. The right column represents the stable structures of the database before adding noise.}
    \label{fig:qualitative}
\end{figure*}
In this experiment, we use our EGNN as a baseline since to the best of our knowledge, there is no equivariant GNN that is capable of acting on both atomic positions and lattice shape at the same time. EGNN can be seen as a hybridisation between a model which only allows acting on the geometry (lattice shape) in an equivariant way (e.g.\ \cite{DBLP:conf/aaai/klipfelPHFSB23}) and a model which only allows acting on the atomic positions (e.g.\ \cite{satorras2021en}).  As can be seen in Table \ref{table:benchmark}, our two models achieve good performances and are able to find the original geometry of the structures in the database. Notice that GemsNet outperforms EGNN for both atomic positions and lattice parameters.


For GemsNet, the reconstruction error on the Perov-5 database is 61\% lower for angles and 20\% lower for the lengths of the sides of the lattice. The positioning error is greater for the atomic positions, but the positioning metrics remain low in both cases.  Recall that Perov-5 is a perovskite base where structures have very similar geometry. As such, it is the least difficult and least interesting dataset. For Carbon-24 the structures are all in pure carbon, but there is more variety in the geometry. GemsNet allows a reconstruction error reduction of 28\% for the lattice parameters and a 20\% reduction in the reconstruction error of the atomic positions. Finally, Mp-20 is the most representative and varied database. On this database, GemsNet allows an average reduction of 10\% of the error of reconstruction of the angular parameters of the lattice; a reduction of the error of almost 40\% of the parameters of lengths and a reduction of 13\% of atomic position error. Finally, we can see that the accuracy of the reconstruction is still very good, especially for the databases which are the most natural, namely with Mp-20 which is a subset of Materials Project.

\subsubsection{Qualitative analysis}


Figure \ref{fig:qualitative} depicts some examples of structures from Materials Project Mp-20 that have been denoised by GemsNet. We can see that the initial structures are too noisy and it is not always easy to reach the stability. One can notice that the majority of the structures are very well reconstructed by GemsNet. Indeed, one can sometimes notice minimal differences in the alignments of the atoms. We can also notice that the shape of the lattices is very close. It is important to note that even crystal structures which are composed of 4 elements are generally well reconstructed whereas many generative models in the literature are limited to 2 or 3 different elements.

However, there are some structures which are visibly poorly reconstructed as we can see in the top right corner of Figure \ref{fig:qualitative}. Several elements are remarkable in these structures. The atomic positions look realistic compared to the stable structures, but it is the lattice that makes the shape of the crystals different. Moreover, the reconstruction error is generally big. Errors produced by GemsNet are therefore rare but they are much more important when they occur. The reconstruction error could therefore come in large part from crystals which are outliers. Indeed, if we take a closer look at the badly reconstructed structure at the top right of Figure \ref{fig:qualitative}, we can notice that this structure is composed of 4 elements. It is therefore a difficult structure to study. Generative models generally struggle to deal with ternary structures (composed of 3 elements) even when they contain a light element such as oxygen. Our structure contains 4 metals including a heavy atom of potentially radioactive uranium, which is uncommon compared to the other elements in the Mp-20  dataset. We can see that GemsNet manages to solve the denoising task when there is uranium in some cases. Moreover, GemsNet handles structures with a few different chemical species, e.g.\ the binary structure at the bottom left of Figure \ref{fig:qualitative}. In general, crystals with heavy atoms are notoriously more difficult to process but GemsNet manages to process some of them. Overall, GemsNet handles a large part of the dataset structures very well but fails on some outliers. These failures can be nuanced because some structures are really more difficult to analyse, even with ab-initio simulation.

\section{Conclusion}


We have proposed two unified equivariant models which act simultaneously on the lattice of a crystal and atomic positions. This is essential for generating new crystal materials since our models are capable to modify the whole geometry during inference and reach the optimum state. While our models provide good performances, GemsNet allows a high level of accuracy for the predicted lattice parameters and for the generated atomic positions. In addition, GemsNet is faster in inference time and contains more than 2 times fewer parameters than existing models. This equivariant architecture opens the way for other applications such as the generation of stable crystalline structures which is a challenging problem.

\section*{Acknowledgments}
This work has been supported by ANR-22-CE23-0002 ERIANA, ANR-20-THIA-0004 and by HPC resources from GENCI-IDRIS (Grant 2022-[AD011013338]).

\bibliographystyle{named}
\bibliography{ijcai23}

\end{document}